\documentstyle[a4,12pt]{article}
\begin{document}
\begin{titlepage}

\vskip 0.4truecm

\begin{center}
{
\bf \large \bf BBGKY hierarchy in scalar QFT. \\
}
\end{center}

\vskip 2.8cm

\begin{center}
{\bf K. A. Lyakhov$^{\sharp}$}\\
\vskip 0.4cm

{\it Sub-department of Quantum Statistics and
Field Theory of  Physical deparment, Moscow State University, 119899 Vorobievy
Gory, Moscow,Russia \\}
\end{center}
keywords:{\it BBGKY hierarchy, Wigner's function, renormalization, scalar
field}.\\
PACS:\\
\vskip 0.7cm
\rm
\noindent
This work is dedicated to obtaining of analog of Bogoliubov's chain for
the case of complex scalar field in QFT and
renormalization problem of obtained equations is discussed.
\vfill

\begin{flushleft}
\rule{5.1 in}{.007 in} \\
\vskip 0.3cm
$^{\sharp}$ \hskip 0.2cm {\small  E-mail: \scriptsize
\bf liakhov4@mail.ru } \\
\end{flushleft}

\end{titlepage}

{\bf Introduction and sculpting the problem.}

In this work attempt to
generalise Bogoliubov's chain,\cite{bogol}, on relativistic case, when the particles number
is not fixed, proceeding from the QFT dynamic equations is perfomed. Our
consideration will be based on the Wigner's function formalism.

{\bf 1.Notations and preliminaries.}

In d-dimensional quantum mechanics, if the system is in a pure state with
the wave function $\Psi$, the Wigner's function looks as
$$
{\cal F}(k,x)=\frac1{(2\pi\hbar)^d}\int dve^{\frac{i}{\hbar}+kv}\Psi^*(x+\frac12v)\Psi(x-\frac12v).
$$
In relativistic quantum statistical physics
this formula transformes as ,\cite{groot}:
$$
{\cal F}(k,x)=\frac1{(2\pi\hbar)^d}\int dve^{\frac{i}{\hbar}kv}
<\varphi^{\dagger}(x+\frac12v)\varphi(x-\frac12v)>,
\eqno(1.1)
$$
where $\varphi^{\dagger}$ and $\varphi$ are the secondary-quantizated field
operators and averaging is performed by Yuttner-Singh distribution, see below.
According to (1.1) the one-particle Wigner's function is the average of
filling number, then the norm condition is:
$$
\int{\cal F}(k,x)dkdx=N
\eqno(1.2)
$$

Let's note through $z_i= (k_i, x_i)$ the coordinate of i-th particle in
relativistic quantum-mechanical phase space.

Let's inductionally introduce n-particles Wigner's function in relativistic case:
$$
\cal F_{\rm n}(\rm z_1\ldots z_n)=N'\int\limits\exp
\left(i\sum\limits_{i=1}^nk_iv_i\right)W_{2n}\left(x_1+\frac{v_1}2,\ldots,x_n-
\frac{v_n}2 \right)dv_1\ldots dv_n.
\eqno(1.3)
$$
as the Furier-transform of the normal-ordered and averaged operators product:
$$
W_{2n}\left(x_1+\frac{v_1}2,\ldots,x_n-\frac{v_n}2 \right)\stackrel{Def}{=}
$$
$$
\stackrel{Def}{=}<:\varphi^\dagger\left(x_1+\frac{v_1}2
\right)\ldots\varphi^\dagger\left(x_n+\frac{v_n}2
\right)\varphi\left(x_1-\frac{v_1}2\right)\ldots\varphi\left(x_n-\frac{v_n}2
\right):>,
\eqno(1.4)
$$
were
$$
N'=\frac1{(2\pi)^{dn}}.
$$
Heretheto we imply
$\hbar=c=k_B=1$ and
field operator $\hat\varphi(x)$ is taken in
Haizenberg representation:
$$
\varphi(x)=U(t)^{\dagger}T(\varphi_0(x)U(t))
\eqno (1.5)
$$
,where $\varphi_0 (x)$ is the solution of the uniform Klein-Gordon's
equation and U(t) is the evolution operator:
$$
U=T exp(i\int\limits_{-\infty}^{t}H_{int}(t)dt)
\eqno(1.6)
$$
Averaging-out in (1.2) is produced as to ordinary rules of
statistical mechanics:
$$
<:O:>=Tr(\rho:O:).
\eqno (1.7)
$$
Here trace is calculated as to any full set of states.
The statistical operator $\rho$ is
the relativistic
analog of the operator realisation of the grand canonical
Gibbs distribution, where the fact, that the particles can produced
and disappear, is allowed for by Haizenberg's representation, see(1.3):
$$
\rho(t)=\frac {1}{\zeta}exp(\beta(\mu N(t)-U_\nu P^\nu(t)),
\eqno(1.8)
$$
where $N(t)$ is operator of number of particles
in Haizenberg's representation:
$$
N (t)=\int d^3p [a^\dagger (\vec p, t) a (\vec p, t) +b^\dagger (\vec p, t) b (\vec p, t)],
\eqno (1.9)
$$
where
$\beta=\frac1{T}$, $\mu$ is the chemical potential; $U_\nu$ is the
hydrodynamical velocity
\cite{groot}, and ${\zeta}$ is defined by the norm condition:
$Tr(\rho(t))=1$.

The 4-momentum operator of the system is
$$
P^\nu (t)=(H (t), P^k (t))
$$
,where $k=1,2,3 $ and $$H(t)=\int d^3pp^0[a^\dagger(\vec p, t)a(\vec p,t)+b^\dagger(\vec p, t)b(\vec p,t)]
\eqno (1.10)
$$
, where $p^0=\sqrt{\vec p^2+m^2}$. Here $a^\dagger$
and $b^\dagger$ are operators of particles and antiparticles production
in Haizenberg's representation correspondingly, see(1.5).

Amenably (1.2) the norm condition for s-
particle Wigner's function takes form:
$$
\int{\cal F}_s(k_1, x_1\ldots x_s, k_s) dz_1\ldots dz_s=N^s,
\eqno (1.11)
$$
where $N$
is the general particles number in the system.

{\bf 2.The model description.}

In this work we'll consider the quantum statistical relativistic system,
consisting from elementary particles, interacting by
Lagrangian of self-interacting complex scalar field:
$$
\cal L=\rm (\partial_\mu\varphi^\dagger\partial_\mu\varphi) m^2
(\varphi^\dagger\varphi)+\lambda (\varphi^\dagger\varphi) ^2
\eqno (2.1)
$$
On the strength of corresponding this model field equations:
$$
(\Box +m^2)\varphi (x)=\rho (x);
\eqno (2.2)
$$
$$
(\Box+m^2)\varphi^\dagger (x)=\rho^\dagger (x)
\eqno (2.3)
$$
, where
$$\rho(x)=2\lambda (\varphi^\dagger(x)\varphi(x))\varphi(x);
\eqno (2.4)
$$
$$\rho^\dagger=2\lambda (\varphi^\dagger (x)\varphi (x))\varphi^\dagger (x).
\eqno (2.5)
$$
we'll deduce BBGKY hierarchy below.

{\bf 3. The kinetic equation for n-partical Wigner's function.}

The Lorentz-invariant kinetic equation for one-particle Wigner's function,
as to essence, was already obtained in the monography\cite{groot}, and
look as:
$$
k_\mu\partial_\mu{\cal F(\rm z)}=i\frac1{2(2\pi)^4}\int e^{ikv}<\rho(x+\frac{v}2)
\varphi(x-\frac{v}2))>dv+h.c.
\eqno (3.1).
$$

By substitution like-current field $\rho$ (2.5) in the given equation we have:
$$
k_\mu\partial_\mu{\cal F} (z) =i\lambda\frac1{ (2\pi) ^4}\int <:\varphi^\dagger
(x+\frac{v}2)\varphi (x -\frac{v}2))\varphi^\dagger(x+\frac{v}2+\frac{v'}2)
\varphi (x+\frac{v}2-\frac{v'}2)):>\times
$$
$$
\times e^{-ik'v'}e^{-ikv}dk'dvdv'+h. c.
\eqno (3.2)
$$
Here we took into account the integral representation for $\delta$-function:
$$
\delta (v)=\frac1{(2\pi)^4}\int e^{-ikv}dk.
$$
So, we have received the first
equation of engaging equations:
$$
k_\mu\partial_\mu{\cal F}(z) =i\lambda\int{\cal F}_2\left (k, x, k', x+\frac{v}2\right)
e^{-ik'v}dk'dv+h.c.
\eqno (3.3)
$$
The subsequent equations of chain
could be found by induction, i.e. it is valid the next

{\bf Proposition 1}
$$
k_\mu \partial_\mu|_{x_i}\cal F_{\rm n}(\rm z_1,\ldots z_n)
$$
$$=i\lambda\int{\cal
F}_{n+1}(z_1,\ldots z_n,z')e^{-ik'v}dvdk'+h.c.,
\eqno(3.4)
$$
where $z'={x_i-\frac12 v,k'}.$

{\bf 4. The renormalization of the chain equations.}

As we well know, in QFT in the force of it's locality singularity arise
at operator-meaning generalized functions(field operators)
production. For their eliminating one resort to renormalization trick
in order to radiation corrections have not led to
infinite expressions for the couple constant and mass,
then in order to search renormalization of the received chain
it is essentially to answer on the following question: "How to distinguish
renormalized(observed) part of the n-particleal Wigner's function
at transition to the semi-QFT (relativistic thermodynamics)?"

For answer on this question advisably to express an arbitrary
operator in terms of reduced density matrices
describing initial state, then the Hilbert space of the system is defined by
initial states,
which are identical with asymptotical\cite{groot} ones ($t\to-\infty$ for
infinity remote past).

For the first we shall consider, for simplicity, the case without
antiparticles, then the asimptotic pure state is:
$$
|p^n\rangle_{\rm in}=a^\dagger_{\rm in}(p^n)|0\rangle,\quad n=0,1,2,\ldots,
\eqno(4.1)
$$

Here we have used the compact notification:
$p^n$ for set
$p_1^{\rm\mu},p_2^{\rm\mu}
,\ldots,p_n^{\rm\mu}$ and $a^\dagger_{\rm in}(p^n)$ for asimptotic operators
product $a^\dagger_{\rm in}(p_1)\ldots a^\dagger_{\rm in}(p_n)$.

If we cut-off Hamilton operator by volume and tribute the periodic boundary
conditions then we have the discrete energetic spectrum corresponding to
Hamiltonian:
$$
H(t)=\int\limits^{p_{max}}_{p_{min}}d^3pp^0[a^\dagger(\vec
p,t)a(\vec p,t)+b^\dagger(\vec p,t)b(\vec p,t)].
$$
The number of particles is finite and is eigenvalue of operator
$$
N(t)=\int\limits^{p_{max}}_{p_{min}} d^3p[a^\dagger
(\vec p,t)a(\vec p,t)+b^\dagger(\vec p,t)b(\vec p,t)].
$$
According to that the volume of our system is very small quantum effects
play essential role. Then energy of system as due to Haizenberg's ambiguity
principle is very large and upper integration limit as to momentum $p_{max}$
could be entered
as particle momentum, generated after interaction immediately. The lower
integration limit
$p_{min}$ could be entered as particle momentum in the time moment of system
evolution accordihg to Haizenberg's ambiguity principle:
$E\sim\frac{\hbar}{\tau}$, where $\tau$ is the life-living time of the emmited
secondary particles.
We stress here, that in this situation {\underline statistical
considerations}
are applied, i.e. the number of particles is large, but finite,
unlike QFT for which, as we well know,
infinitely large degrees of the freedom (field quants) is characteristic.
In force that the number of particles is finite the completeness condition
is next:
$$
{\bf 1}=\sum\limits_{n=0}^{N(E)}\frac1{n!}\int\limits
\frac{d^3p^n}{p^{0n}}|p^n\rangle_{\rm in}\;_{\rm in}\langle p^n|,
\eqno(4.2)
$$
where N(E) is the function of energy of the system. Then the modification of
formula for average for arbitrary
operator $O$, ref.\cite{groot}, with take into account new
completeness condition(4.3)is neaded:
$$
\langle O \rangle=\sum\limits_{n=0}^{N(E)}\frac1{n!}\int\limits
d^4x^nd^4k^nO_n(x^n,k^n)\sum\limits_{m=0}^{N(E)}\frac{(-1)^m}{(m!)^2}\int\limits
d^4x^md^4k^m\prod\limits_{j=1}^{n+m}{\cal F}_{\rm
in}(x_j,k_j).
\eqno(4.3)
$$
Here the integrals as to momenta are cutted off on upper and lower
limits according with above said and
integrals as to space are cutted off in accordance with the one meson space volume.
Multiplier $\sum\limits_{m=1}^{N(E)}\frac{(-1)^m}{(m!)^2}\int\limits
d^4k^md^4x^m\prod\limits_{j=1}^{m}{\cal F}_{\rm
in}(x_j,k_j)$ equals
$$
Z=-<N_{in}(E)>+\frac14(<N_{in}(E)>)^2-\frac1{36}(<N_{in}(E)>)^3+\ldots+\frac{(-
1)^{N(E)}}{(N(E)!)^2}
(<N_{in}(E)>)^{N(E)},
\eqno(4.4)
$$
where we used, that
$$
\int{\cal F}_{in}(k, x)dxdk=<N_{in}>
\eqno(4.5)
$$.
$N_{in}$ is the number of scalar particles operator in initial
state. Let's examine the
Wigner's function expansion as to the contributions of collisions. After
substitution (4.5) into (4.4) we have:
$$
\langle O \rangle=\sum\limits_{n=0}^{N(E)}\frac1{n!}\int\limits
d^4x^nd^4k^nO_n(x^n,k^n)\prod\limits_{j=1}^n{\cal F}_{\rm
in}(x_j,k_j)\times Z
\eqno(4.6)
$$
Having applicated (4.6) to the one-particle Wigner's function, we have
the following expansion:
$$
\cal F(\rm x,k)=\sum\limits_{n=0}^{N(E)}\frac 1{n!}\int d^4x^nd^4p^n
\Psi_n(x^n,p^n|k,x)\prod\limits_{j=1}^n\cal F\rm _{in}(x+x_j,p_j)Z
\eqno(4.7)
$$
,where
$$
\Psi_n(x^n,p^n|k)=N'\int d^4u^n\prod\limits_{j=1}^{n}e^{iu_jx_j}{} _{in}
\langle p^n-\frac 1{2}u^n|\Psi(k)|p^n+\frac 1{2}u^n\rangle_{in}.
\eqno(4.8)
$$
,and operator $\Psi(k)$ is:
$$
\Psi(k)=\frac1{(2\pi)^4}\int d^4ve^{-ikv}\varphi^\dagger\left(\frac12
v\right)\varphi\left(-\frac12 v\right).
\eqno(4.9)
$$

Let us calculate every matrix element $\Psi_n(x^n,p^n|k)$ separately.
Matrix element equals zero at $n=0$, as it have to be for normal-ordered operator.

We have used the completeness condition (3.7) at $n=1$, which leads to
the following expression:
$$
{\rm\Psi}_1(x,p|k)=\frac1{(2\rm\pi)^4}\sum\limits_{m=0}^{N(E)}
\frac1{m!}\int\frac{d^3p^{\prime m}}{p^{\prime 0m}}
 {\rm \delta}^{(4)}
\left(k+\sum\limits_{j=1}^mp'_j-p \right)\times
$$
$$
\times\int d^4u\exp(iux)_{\rm in}\Bigl\langle p-\frac12u|{\rm
\varphi}^\dagger(0)|p^{\prime m})\rangle_{\rm out}\;_{\rm out}\langle
p^{\prime m}|{\rm \varphi}(0)\Bigr|p+\frac12u\Bigr\rangle_{\rm in}.
\eqno(4.10)
$$
Matrix elements $\Bigl\langle p-\frac12u|{\rm
\varphi}^\dagger(0)|p^{\prime m}\rangle_{\rm out}$ can be calculated by
using Jang-Feldman equation.
$$
\langle0|{\rm \varphi(0)}|p \rangle_{\rm in}=
\frac1{\sqrt{2p^0}}\frac1{(2\rm\pi)^{3/2}}.
\eqno(4.11)
$$
Finally we have
$$
{\rm\Psi}_1(x,p|k)={\rm \delta}^{(4)}(k-p){\rm \delta}^{(4)}(x).
\eqno(4.12)
$$
{\bf Proposition 2}

Using the results (4.10)-(4.12), the expansion of
Wigner's function  as to the contributions of collisions is:
$$
\displaystyle{\cal F}(\rm x,k)=Z\times[{\cal F}_{\rm
in}(\rm x,k)+\sum\limits_{n=2}^{N(E)}\frac1{n!}\int
d^4x^nd^4p^n{\rm\Psi}_n(x^n,p^n|k)\times
$$
$$
\times\prod\limits_{j=1}^n\cal F_{\rm in}^{(+)}(\rm x+x_j,p_j)].
\eqno(4.13)
$$
It is impossible to overlook obvious similarity received expression
with virial expansion for the classical distribution function in
nonrelativistic statistical physics.

{\it Remark:}
According to above mentioned assumption about
initial state of the system, we see, that the first term
corresponds to solution of uniform or colisionless kinetic
equation:
$$
k_\mu \partial_\mu {\cal F}_{in}(z)=0.
\eqno(4.14)
$$
Initial distribution
function is the equilibrium relativistic Boze-Einstein distribution function as
it discussed above, but without spin:
$$
{\cal F_{\rm in}}(k,x)=\frac1{(2\pi)^3[e^{\beta(k_\nu U^\nu(x)-\mu(x))}-1]}
\eqno(4.15)
$$
The rest of the members allow for the contribution of interparticles
interactions. According to (4.5) $Z$ is the finite value unlike QFT, where
it would be diverging quantity and accords to renormalization of wave function:
$\varphi^{R}(x)=Z_\varphi^{-\frac12}\varphi(x)$, see\cite{shirk} due to
Dyson's transformation.
Thus the following statement is valid:

{\bf Proposition 3}:
$$
\left. k_\mu^n \partial_\mu \right|_{x_i}F_{n}(z_1 \ldots z_n)=
$$
$$
= iN'Z\lambda\int\limits_{k'_{min}}^{k'_{max}} F_{n+1}(z_1 \ldots z_n,z')e^{-
ik'v}dvdk'+h.c.
\eqno(4.16)
$$

{\bf 5. The physical illustration.}

Received chain may be applicated for the computing of kinetic
coefficients in non-linear approach, i.e. when the effect of
correlations between particles is essential. By other words,it is
necessary to allow for not only interaction of particles with each
other, but also the interaction of particles with clusters. For example,
such situation could be emerged in description of phase transition in
relativistic plasma. The role of
clusters play numerous short-living resonances in the elementary particles
physics. Conjecturally adequate
mathematical tool for non-equilibrium
interactions between clusters (resonances) description is
the Bogoliubov's chains
method(BBGKY).
Let's for definity consider the following many body processes:
$$
pp\to pp\pi^+\pi^+\pi^-\pi^-\pi^0\to p\Delta^{++}\pi^+\pi^-\pi^-\pi^0
$$
$$
pp\to pn\pi^+\pi^+\pi^+\pi^-\pi^-\pi^-\to p\Delta^+\pi^+\pi^+\pi^-\pi^-\pi^-
\eqno(5.1)
$$
and etc.
The volume, within that collision occurs and the number of generating
particles could be calculated, for example, from multiperipheral model,
\cite{pignotti}. In this
article it was built the quantitative theory of the multiple
mesons production with respect of secondary particles interactions at
ultra-high energies. There it was
obtained, what the number of particles logarithmicaly depends on energy $E$
generated by collision:
$N(E)\sim ln s$, where $E\sim\sqrt{s}$.
Respected that volume of system is Lorentz-contracted
$V\sim\frac1{m_{\pi}^3}\frac{2M}{E}$, where M is nucleon mass, maximal
$\pi$-meson energy-momentum equals
$$
p_{max}^{\mu}=(p_{max}^{0}=\sqrt{\vec p_{max}^2+m^2}\sim\frac{E}{N}\sim
\frac{ln s}{\sqrt{s}},\vec p_{max}).
\eqno(5.2)
$$
In the case of nucleon beams collision all emitted energy is
exuded in the small volume and spent to pionic stars formation and
belated to the binded states formation or resonances, see.(5.1). With the
flow of time the system expands and becomes more dilute.
I.e. we have a number of, step by step alternating each other
descriptions: local or QFT-description, then thermodynamical follows, which
replies to nucleon plasma on the early stages of it's formation corresponding
to the high energies, no more than 10GeV or hydrodynamical description \cite
{landau} for
ultra-high energies, then kinetic stage follows, at which
correlations accord essential influence on the system dynamics and eventually
passes in equilibrium state, corresponding to Yuttner-Singh distribution.

{\bf Summary.}

The kinetic stage of the evolution of the system of scalar particles
interacting as $\lambda(\varphi^\dagger\varphi)^2,$ is
described by the chain of equations (4.16).

\end{document}